# Obtaining physical layer data of latest generation networks for investigating adversary attacks

**M.V. Ushakova, Yu. A. Ushakov, L.V. Legashev**
Russia, Orenburg, Orenburg State University

e-mail: m.v.ushakova@mail.ru

**Abstract**. The field of machine learning is developing rapidly and is being used in various fields of science and technology. In this way, machine learning can be used to optimize the functions of latest generation data networks such as 5G and 6G. This also applies to functions at a lower level. A feature of the use of machine learning in the radio path for targeted radiation generation in modern ultra-massive MIMO, reconfigurable intelligent interfaces and other technologies is the complex acquisition and processing of data from the physical layer. Additionally, adversarial measures that manipulate the behaviour of intelligent machine learning models are becoming a major concern, as many machine learning models are sensitive to incorrect input data. To obtain data on attacks directly from processing service information, a simulation model is proposed that works in conjunction with machine learning applications.

## 1. Introduction

Modern artificial intelligence methods are actively used in solving network security problems, namely, classifying network threats based on analysis of network traffic and identifying malicious network activity features. It is worth noting that the artificial intelligence models themselves are also subject to various types of network attacks with the aim of maliciously compromising the output data.

There are two main scenarios for attacks on machine learning models:

- “white box” attacks are typical for the case when an attacker has direct access to a machine learning model with the ability to study the source code, architecture, masking and strengthening of individual features;
- “black box” attacks are typical for the case when an attacker does not have direct access to a machine learning model, but has the opportunity to test the model on specially prepared malicious data to identify its weaknesses.

An attack on a neural network and/or a machine learning model is called an adversarial attack and its goal is to deliberately distort the output data when specially prepared (poisoned) data (so-called adversarial samples) is supplied as input data. The attacker's goal in this case is to reduce the efficiency of trained machine learning models or to bring the output values closer to a pre-calculated incorrect result.

If 2/3/4G networks were built mainly on the basis of hardware solutions (equipment), then the 5G platform (and in the future 6G) is built on the basis of software solutions, in particular, software-defined networks (SDN), as well as of network functions virtualization (NFV), and machine learning models are responsible for automatic configuration of parameters of services, applications and scaling [1, 2, 3].

The authors of papers [4] and [5] show the possibilities of attacks on 6G networks using adversarial networks. Articles [6], [7] show the relevance of researching such adversarial attacks on physical infrastructure, and new types of antenna and radio-photonic equipment for 6G networks will initially rely on machine learning methods for beamforming, which increases the need for research on protection against attacks on infrastructure.

## 2. Theory

The latest generation networks - 5G and 6G - require increasingly flexible approaches and technologies for the use of radio equipment and new principles for the formation of high-speed directed data transmission. Traditionally, radio link capacity has been increased through multiplexing of space, frequency, time, coding and polarization. Then, to improve data transmission characteristics, a multi-beam MIMO antenna with different polarization, new modulation schemes, and new signal shapes was used. The use of multipath data transmission may cause fluctuations in amplitude, phase and angle of arrival, resulting in a fading effect. The physical cause of fading is, in addition to multipath propagation, mainly changes in the parameters of the propagation medium (inhomogeneity), reflections generating echo signals, the presence of reactive elements in the transmission path, and possible Doppler effects. All 4G and 5G standards use OFDM as the main signal modulation to combat multi-path fading, but the latest generation networks require higher frequencies to provide higher throughput, where older methods have low efficiency.

Spectrum in the terahertz range - sub-terahertz and millimeter bands - will be the base spectrum in 6G cellular networks, the lower THz band (0.3–1.0 THz) will be the main candidate for short- and medium-range data transmission. The physical layer for organizing data transmission in these ranges differs significantly from the traditional 1-6 GHz ranges; even for the millimeter range 5G mmWave (26-28 GHz, 39 GHz) already directed beams and an active multi-element antenna array are required. The sub-terahertz range differs significantly from the millimeter band in terms of channel characteristics, device design, signal generation and antenna technology. 5G networks use massive-MIMO, in which the number of antennas must be greater than the number of end users, 6G networks will use even more antennas, as a result of which it is necessary to more accurately obtain the characteristics of wireless channels to adjust the antenna and channel state information matrices (CSI), on the basis of which the beam is formed (see figure 1).

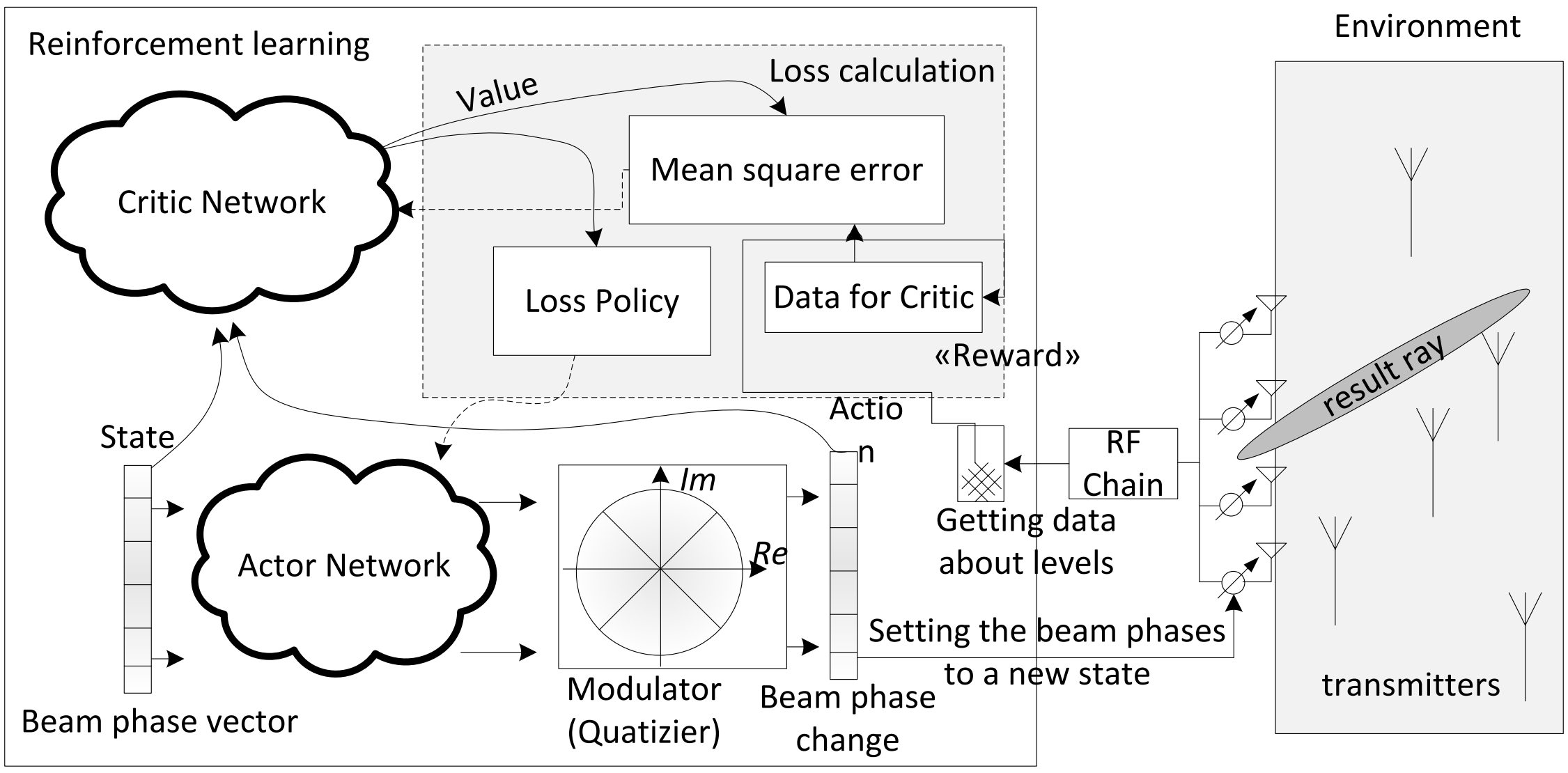

Figure 1 – The principle of narrow beam formation using machine learning.

To analyze possible paths of adversarial attacks on the physical layer, a classification of the application of artificial intelligence and machine learning methods to components of the physical layer was carried out. As a result, the following main areas of use at the physical level were identified:

1) studying the property of temporal correlation of the wireless channel for proactive control of parameters;

2) neural network as a channel decoder, joint source and channel coding, modulation and demodulation;

3) direct selection of channel parameters for proactive beamforming;

4) determining modulation and coding settings for proactive control of parameters based on feedback from receivers.

Since the problem of generating optimal beam pairs together with antenna tuning problem is an NP-complete problem, most smart tuning tools use a hybrid approach to solve it, in which a deep feedforward neural network helps the traditional computational method optimize the parameters. However, for 6G generation networks, the use of both orthogonal and non-orthogonal code division multiple access methods is predicted; therefore, the range of neural network technologies in terms of their influence on beamforming is considered together with the formation of codewords and channel state matrices (formation of the generative matrix of the spatial code).

Attack on machine learning models through influence on the physical data received by base stations and radio transmitting equipment (fake device reports, influence on the signal-to-noise ratio of the reception, deliberate changes in signals over time, dropouts, calculated modulation of transmitter power, exploitation software-controlled surfaces) although it is not an attack through third-party channels (side-channel attack), but is close to them in its idea. The formation of the codebook for the antenna array is shown in figure 2.

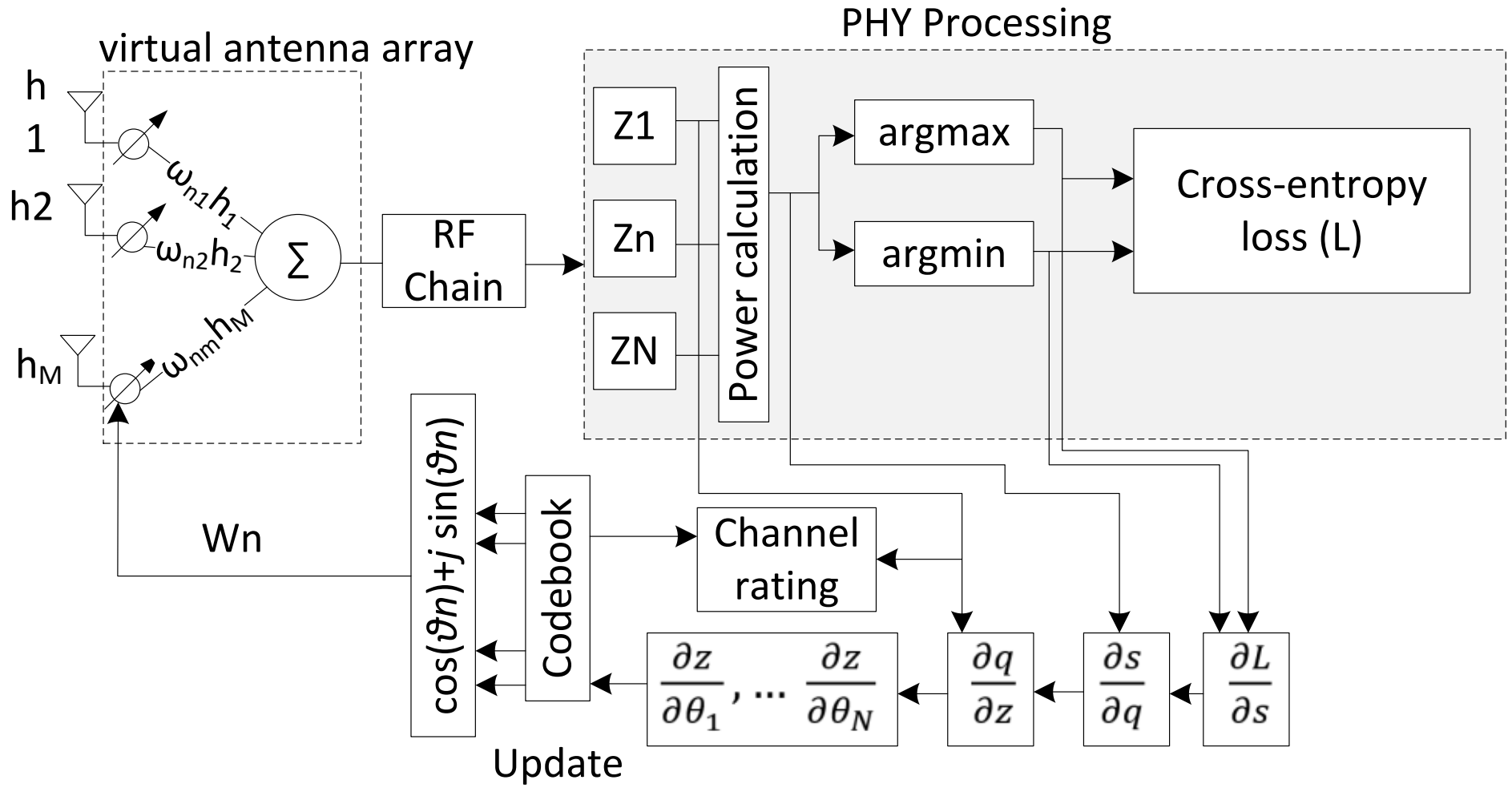

Figure 2 – Classic brainforming.

Since it is necessary to track the consequences of an attack both retrospectively and in real time, it is necessary to organize a system for collecting data on all possible metrics and processing them in real time. Based on the research of the 6G-SANDBOX consortium [8], which is developing an experimental 6G ecosystem on top of existing 5G technologies and new developments, an open source emulator was taken as a basis to study the problem of attacks on 6G infrastructure OMNet++ 6.0 [9] (with the INET 4.5 framework), the radio interface level simulator Simu5G 1.2.2 [10] and the dataset generator for DeepMIMO 5GNR radio interfaces together with a calibrating data set for the mmWave range [11]. Since this work only considers the physical layer (air interfaces), emulation of the 5G core is not required. Based on these tools, a simulation of the radio part has been launched, which makes it possible to obtain information about events and actions on the part of the radio interfaces. To connect the simulator and a real application, the OpenNESS 20.06 framework was used, which allows you to connect real applications to the simulation, implementing the Multi-access Edge Computing (MEC) environment.

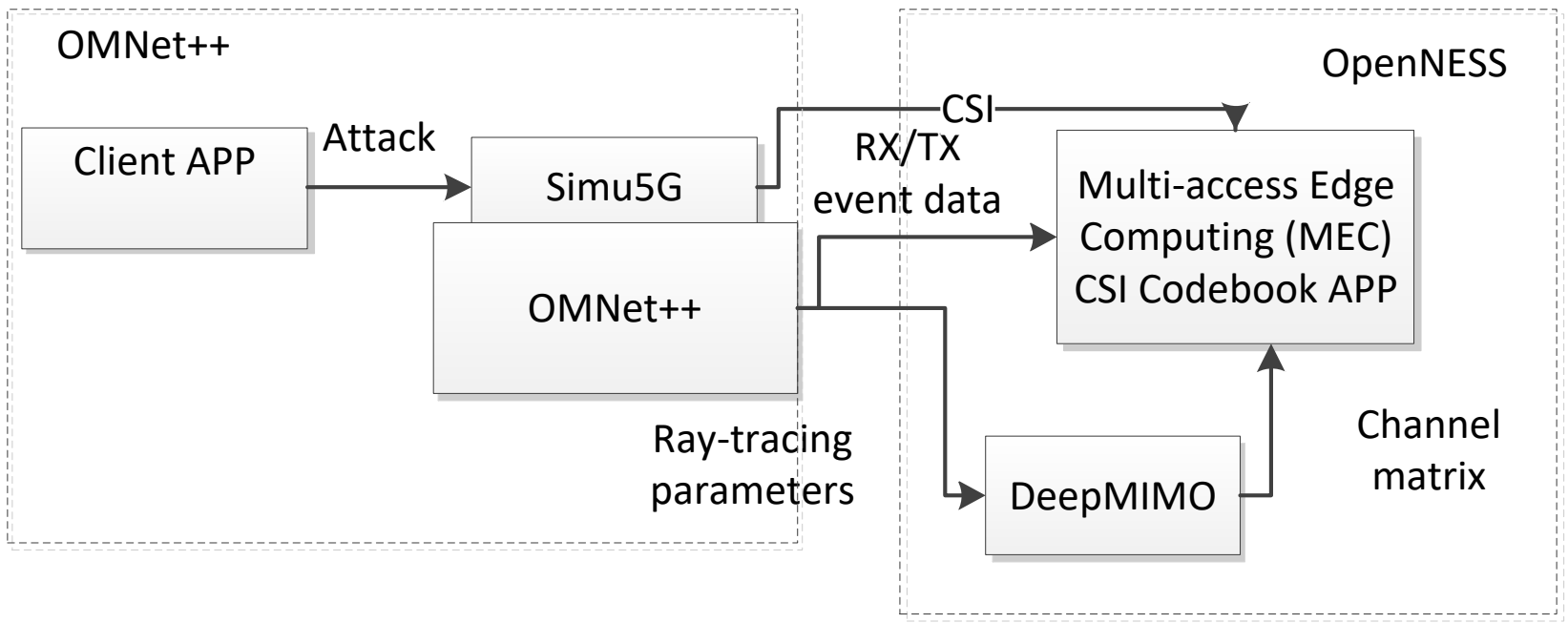

Figure 3 – Modeling tools links.

The study examined and analyzed the following parameters that influence beam formation and are available for external influence:

1 Adaptive modulation and coding (algorithms based on CNN, LSTM, FCNN), the target of the attack is erroneous classification of modulation; implementation of the attack - broadcast modulated radio signal; attack vector - classification type predictor for the decoder.

2 Channel coding (algorithms based on autoencoders, RL), the target of the attack is an error in decoding information; implementation of the attack - broadcast modulated radio signal; attack vector - decoder based on intelligent models.

3 Beam formation (FCNN-based algorithms), the goal of the attack is classification error and selection of a non-optimal beam, maximizing the model error compared to real data; implementation of the attack - broadcast modulated radio signal, attack on RSS and CSI data; attack vector - generating matrix generator of spatial code based on intelligent algorithms.

4 Channel state estimation (algorithms based on CNN autoencoders), the goal of the attack is to maximize the error between real and model CSI; implementation of the attack - broadcast modulated radio signal and falsification of response user data; attack vector - an intelligent model of the decoder, namely a generating matrix, a code word.

5 Signal strength (algorithms based on FCNN, DQN), the purpose of the attack is to exceed the capabilities of the transmitter, reduce the speed; implementation of the attack - broadcast modulated radio signal and position spoofing; attack vector - an intelligent model of the decoder, namely a generating matrix, a code word.

6 Planning and allocation of resources (DQN-based algorithms), the goal of the attack is erroneous classification, and as a result, an increase in the number of errors; implementation of the attack - broadcast modulated radio signal and data falsification, fake requests; attack vector - intelligent planner model.

## 3. Experiment

A description of the implementation of the attacks themselves is beyond the scope of this paper and will be discussed in other relevant publications. DeepMimo 5G NR was used to obtain data on the physical model of beam propagation, as well as data on distances, losses, signal reception and sending angles, phase, power, and signal propagation path. Changes have been made to the reporting formats to model the physics in more detail and obtain parameters such as in-phase and quadrature modulation data, accurate signal-to-noise ratios, modulation and encoding data used for beamforming. Due to the fact that a reconfigurable intelligent surface (RIS) is used in conjunction with traditional phased array antennas, codewords for each type of emitter were studied separately. Based on the shortcomings in the field of instability to adversarial attacks in the ML/AI-based beamforming prediction methods considered by the team (for both line-of-sight and non-line-of-sight cases), attack detection options are considered:

1 Detection of attacks based on testing the hypothesis that in a given block of fading for known channel gains the most powerful signal is maliciously modulating together with the developed method for assessing the stability of the model based on the GLR-CUSUM method. In this case, a dynamic model of the Rayleigh fading channel, a phase shift model and an estimated CSI matrix together with the phase shift vector are formed.

2 Detection of attacks using the CSI statistical error model to model the uncertainty of channel parameters and statistically identify differences compared to the reference distribution of parameters using the Kolmogorov-Smirnov criterion. According to the worst-case scenario of CSI matrix uncertainty, we consider a quasi-static block with flat fading and a continuous phase shift model.

3 Using GAN to detect fake channel state information. In this case, the generative model minimizes the difference between real and fake data, while the discriminative model tries to maximize this difference. Network model D contains four layers consisting of Convolutions, Batch Normalization, Leaky ReLU and Dropout layers. The output layer consists of a linear activation function followed by a mean square error (MSE) layer that acts as a loss function. The inputs to D are the real link state data (H) and the network G generated link state data (set H'). Network G is built as a channel state estimation block with encoding and decoding blocks. It generates H' from the noisy receiver signals (Y). Since the signals received by the 6G base station are combined with noise, the received signals are processed by an autoencoder. Network G consists of four encoding blocks and four decoding blocks. Each encoding block consists of Convolutions, Batch Normalization and ReLU layers, and the decoding block consists of Transposed Convolutions, Batch Normalization and ReLU. Since all attack detection options require data on CSI matrices and data logs affecting it, data collection from all radio interfaces is required.

For prototyping, a model based on the OmNET++ simulation tool was implemented. Three transceivers with weak adapters (frequency 24 GHz, antennas 2 dBi) moving in a circle, emulating vehicles, were used. Each vehicle periodically attacks the matrix by modulating the signal and sending incorrect response information about the state of the channel. Transmitters are periodically attacked by emulating incorrect reception conditions; the vehicles are constantly trying to upload information to the server and receive other information from the server, which is recorded in the log (figure 1).

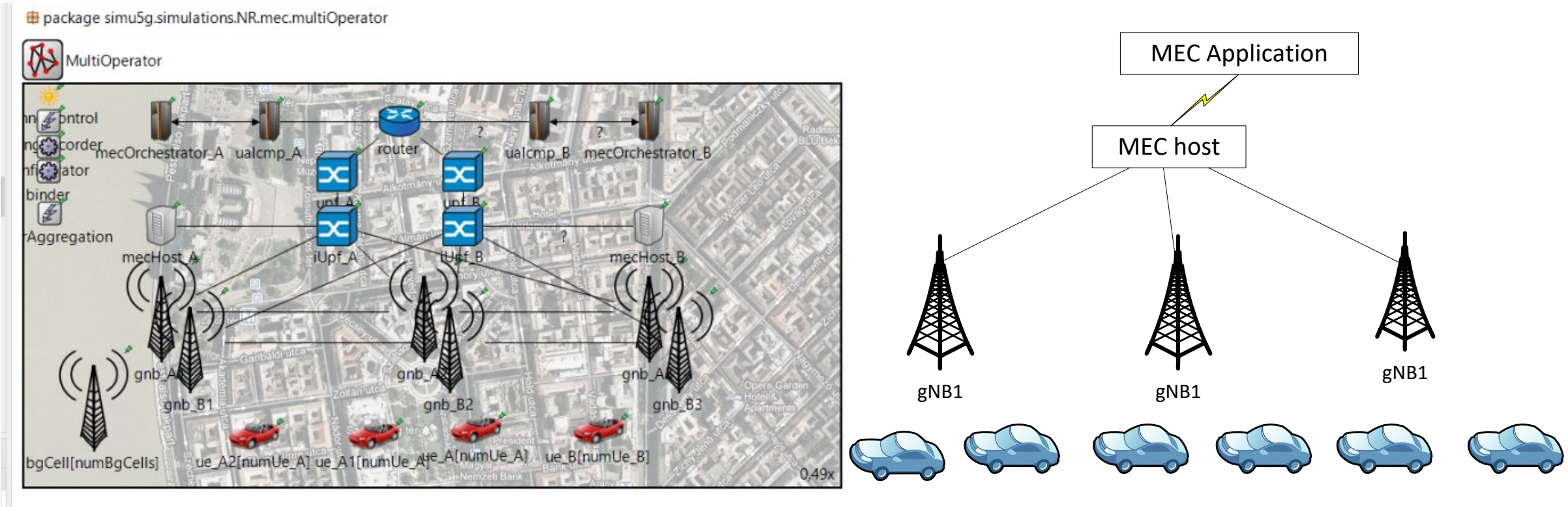

Figure 4 – Modeling window (a) and traffic path (b).

For interaction in conditions of zero trust networks, when launching vCPE for SD-WAN, the address of several Config server addresses is specified, through which updates are downloaded via the established DMVPN channel. A node is used as a base station, which periodically updates information from the Config server, is itself a carrier of one of the Config server addresses and participates in the VPN network (if there is a wireless connection). As a result, vehicles that find themselves within the shielding can only communicate with each other and with the base station. The physical layer models the effects of signal propagation and interference on the MAC layer transport block (rather than on the component bits/symbols). Each block is encapsulated in an AirFrame OMNeT++ message.

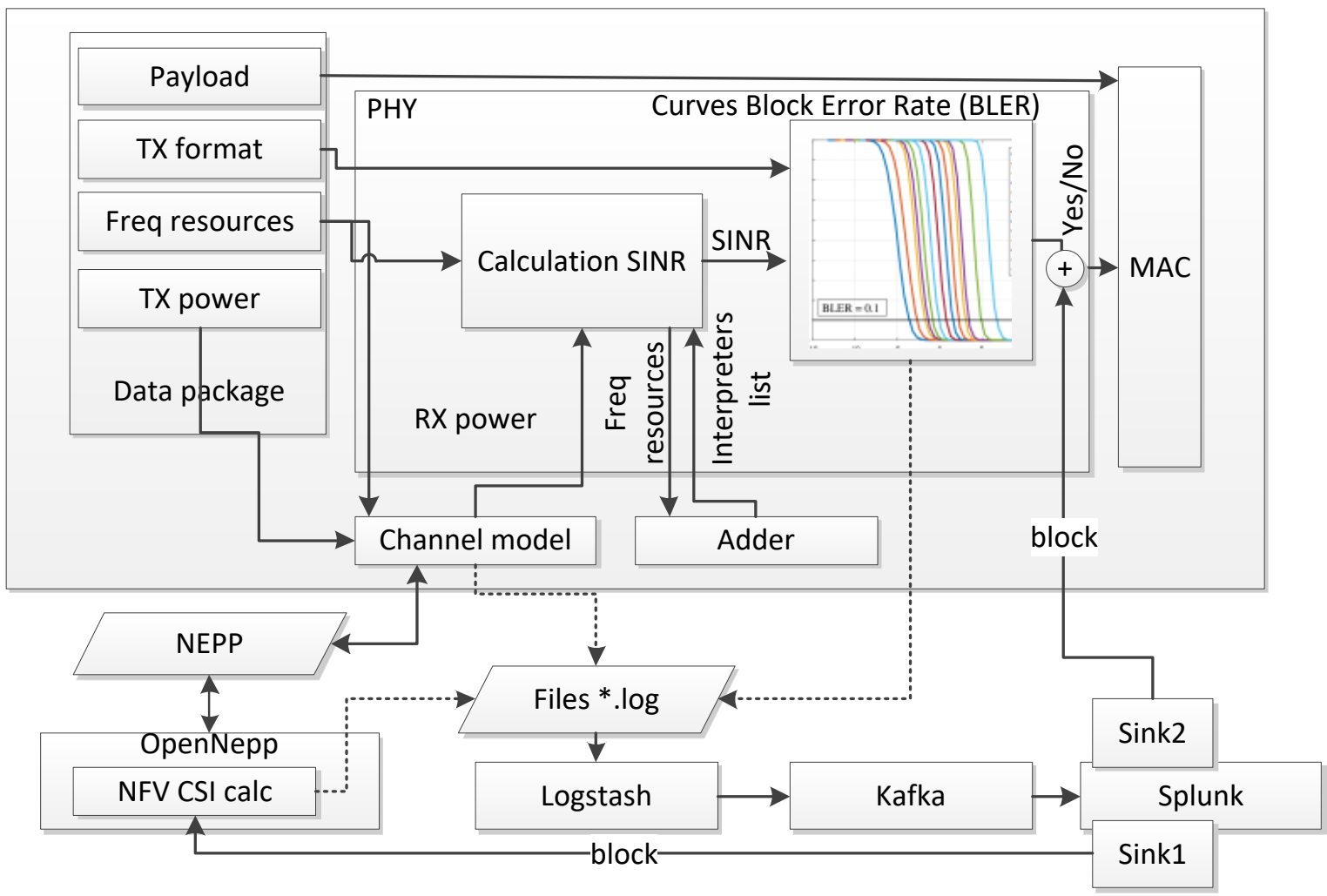

Figure 5 – Data collection scheme at the physical level of the model.

Directions for the use of intelligent methods in beamforming applications using phased arrays and intelligent reconfigurable surfaces have been explored, machine learning and artificial intelligence technologies that can be used to train such systems and control the parameters of radio systems that affect MIMO distribution, spatial coding formation and formation of code words. In such conditions, the percentage drop in throughput depends on the probability of detecting attacks, subject to different values of false positive probability. The latency with enabled detection and with disabled detection is shown in figure 6.

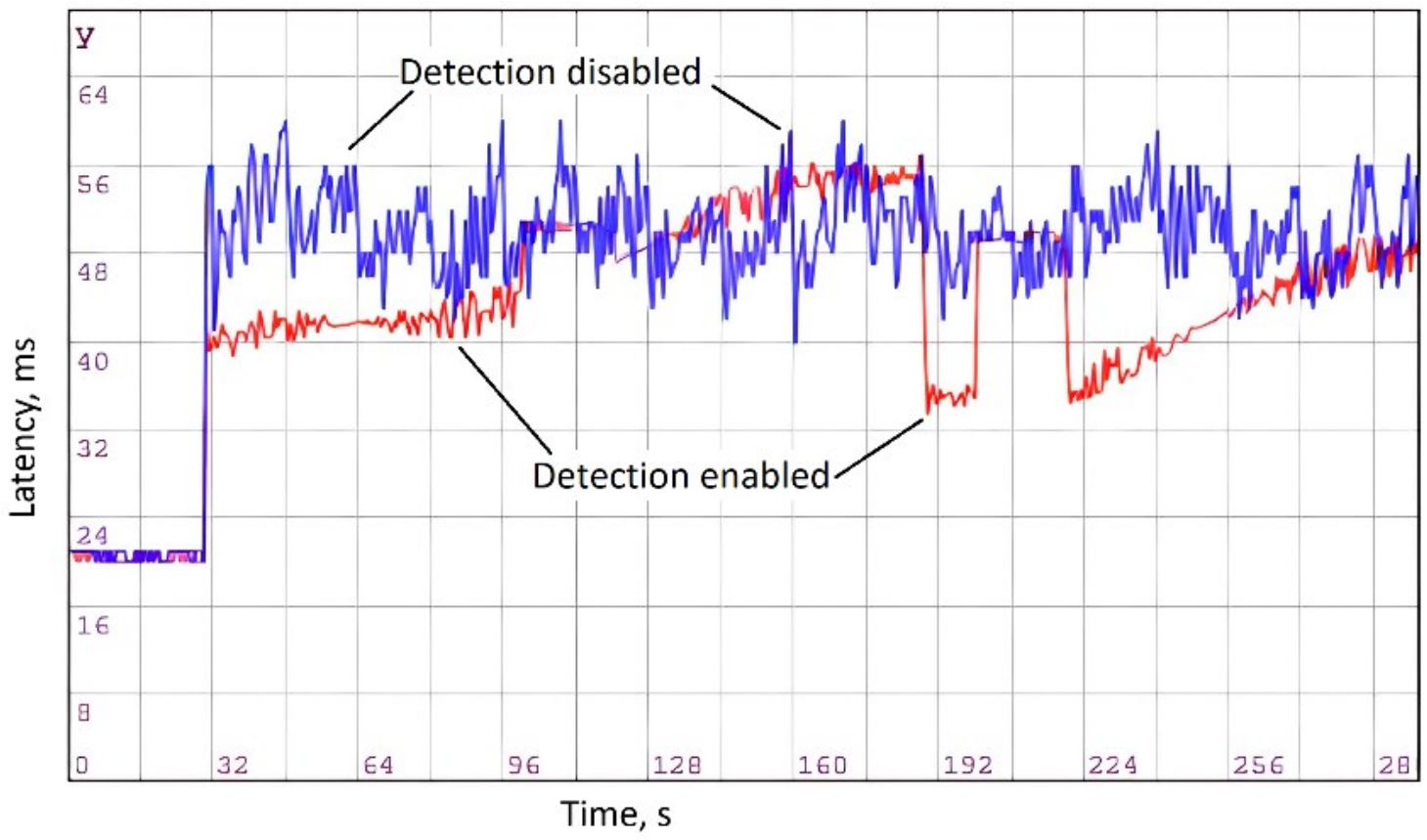

Figure 6 – Simulation results: latency behavior with enabled and disabled attack detection.

Since the base station must act as an intermediate node to encrypt traffic, information about access rights and keys is periodically updated, if the vehicle does not update the information, it will require a long identification procedure, and accordingly there will be either packet loss or large delays of some network traffic in the queue.

As a result of modeling attacks in the DeepMimo system, the transmit power parameters were 10dB, the number of antenna elements was 64, the number of receivers was 50, the data was averaged for 10,000 different channels, the following results were obtained for attack directions:

- With a difference between the signal-to-noise ratio during normal operation and during an attack on coding by 10%, the speed reduction reached 9.3%, for 20% - 17.7%, for 30% - 25.6%, for 40% - 35.9 %. At the same time, the probability of successful detection using method 1 was (at a level of no more than 0.1 false positives) from 0.5 to 0.6, with the number of analyzed data samples from 50 to 100.
- When attacking the CSI matrix (phase component) using a significance level of 0.01 for a false negative probability threshold of 0.1, the algorithm showed the need for at least 120 measurements, while the probability of detecting an attack was 0.49. For a false negative probability threshold of 0.2, the algorithm showed the need for at least 75 measurements, while the probability of detecting an attack was 0.3.

## 4. Conclusion

The main directions of work related to the operation of systems in the millimeter and sub-terahertz ranges are considered, and a review of attacks in these ranges to influence the formation of ultra-narrow beams in the conditions of using neural networks and machine learning methods is also carried out in order to implement massive-MIMO and ultra-massive MIMO technologies together with new coding and modulation methods.

The main directions of research into attacks on the considered intelligent technologies have been formed:

1 Adaptive modulation and coding.
2 Channel coding.
3 Beam formation.
4 Channel condition assessment.
5 Signal strength.
6 Planning and resource allocation.

A simulation system based on OMNeT++ and Simu5G was created, modules were added to the model for obtaining physical layer data, external applications were added to determine the parameters of the radio path during beam formation, and the possibilities of influencing the radio path were studied.

The research was funded by the Russian Science Foundation (project No. 22-71-10124).